\documentclass[aps,apl,twocolumn]{revtex4} 

\usepackage{graphicx}
\usepackage{amssymb}
\usepackage{bm}
\usepackage{xcolor}
\begin{document}

\title{Determining the parameters of a random telegraph signal by digital low pass filtering}
\author{Shilpi Singh, Elsa Mannila, Dmitry Golubev, Joonas Peltonen and Jukka Pekola}
\affiliation{
	 QTF Centre of Excellence, Department of Applied Physics, Aalto University, FI-00076 Aalto, Finland.  
	}

\begin{abstract}
We propose a method to determine the switching rates of a random telegraph signal. We apply digital low pass filtering with varying bandwidth to the raw signal, evaluate the cumulants of the resulting distributions and compare them with the analytical prediction. This technique is useful in case of a slow detector with response time comparable to the time interval between the switching events. We demonstrate the efficiency of this method by analyzing random telegraph signals generated by individual charge tunneling events in  metallic single-electron transistors.
\end{abstract}

\maketitle

Telegraph noise, or random switching of the electric current between two levels,
is often observed in electronic devices \cite{Weissman}. In the context of nanotechnology, it is
particularly important for semiconducting \cite{KU,MOSFET} and single-electron transistors \cite{NA1,SET,Hofheinz}, 
as well as for solid-state quantum bits (qubits) \cite{qubit,Delsing,Ustinov}. Switching rates between the two current levels,
and their dependence on temperature, gate voltage or other parameters, provide 
valuable information about the physical nature of the two-level systems generating the noise.
The standard way of finding these rates is based on threshold detection algorithms, in which
the detector current values  below or above certain threshold are assigned to the first or the second   
state of the two-level system.  

Accurate determination of the switching rates may be hampered by white and $1/f$-noises  
present in the output signal of the detector, and by the long response time of the latter. 
Several methods of correcting the errors caused by these effects have been developed in the past.
For example, the white noise is efficiently suppressed by digital low pass filtering. 
Unfortunately, during this procedure switching events separated by short time 
intervals are lost as well \cite{NA}. In order to take that into account,
Naaman and Aumentado \cite{NA} have introduced two additional states of the system 
corresponding to the errors occurring when the detector does not immediately switch after the jump 
in the telegraph signal, 
and introduced the decay rate of these states as an additional parameter.
Yuzhelevski {\it et al.} \cite{YYJ} have proposed an iterative procedure of the analysis of noisy data,
in which the thresholds for the jumps between the two levels are adjusted based on the
values of the switching rates found in the previous iteration. Martin-Martinez {\it et al.} \cite{Martinez}
have introduced weighted time lag method relying on the analysis of correlations between the
neighboring points in the digitized noisy signal. 
K\"ung {\it et al.} \cite{PhysRevB79035314} 
have proposed a cross-correlation technique with two detectors;
and Prance {\it et al.} have used wavelet edge detection technique \cite{Prance}. 
All these techniques have been tested in practice and proved to be efficient \cite{Ensslin,Haug,Slichter}.
Additional complications arise if several two level systems contribute to the noise and
the current switches between multiple levels. Awano {\it et al} have developed an algorithm,
based on the theory of Markov chains \cite{Rabiner} and Monte Carlo simulations, capable of determining the parameters of
all two level systems \cite{Awano}. Similar approach have been used by Puglisi and Pavan,
who have analyzed noise in random access memories \cite{Puglisi}.
Giusi {\it et al} have proposed an algorithm of separation of the two and multilevel
telegraph noise from background $1/f$-noise \cite{Giusi}. 

\begin{figure}
\includegraphics[width=0.9\columnwidth]{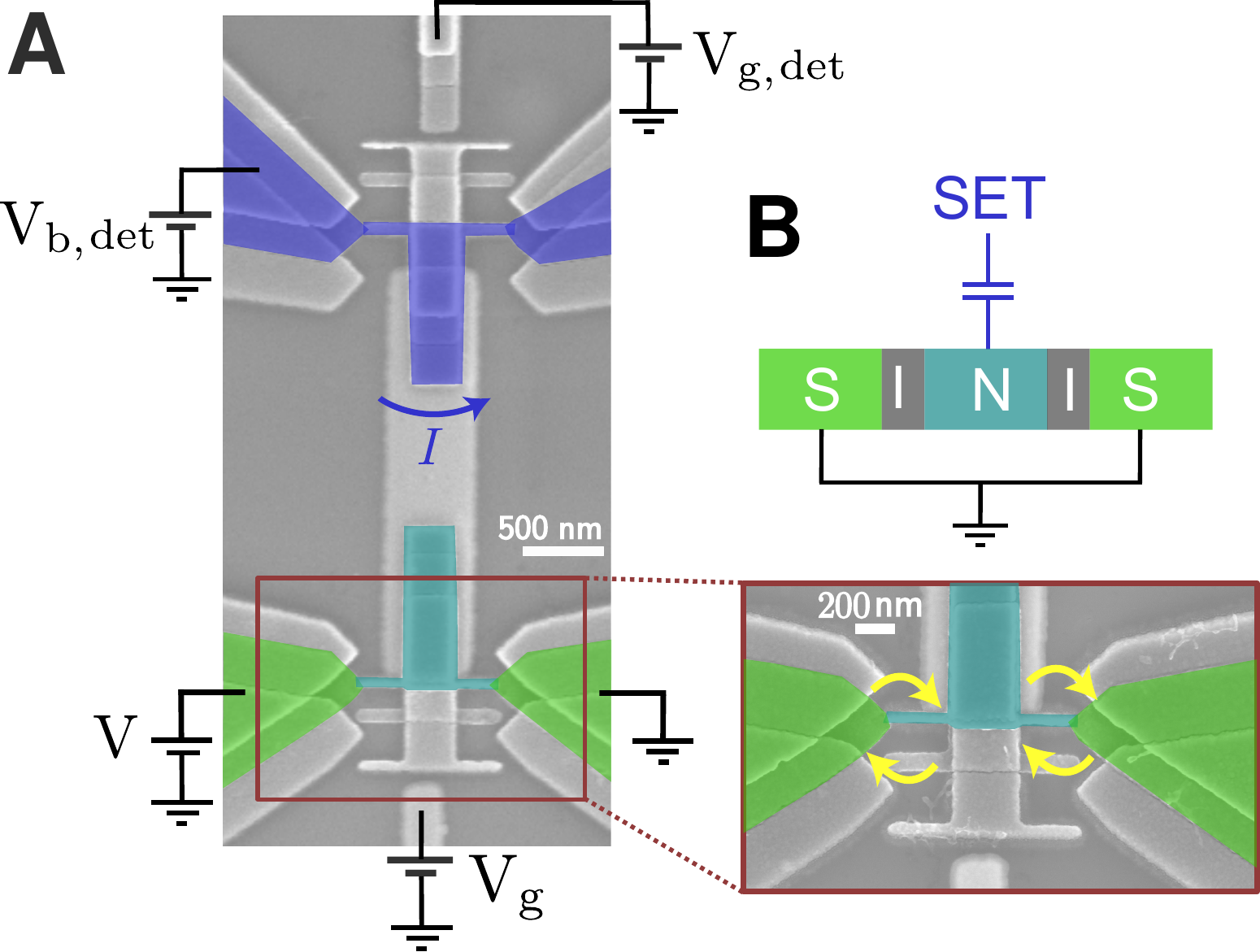}  
\caption{
	\textbf{Experimental setup.}  
	 (a) Pseudo-colored scanning electron micrograph of the detector SET (blue, top) and
	  another SET 
	  used as the noise source. 
	  (b) Top: schematic sketch of the measurement setup. Bottom: Zoomed view of the lower SET, which has two superconducting leads (S, green) and normal-metallic island (N, turquoise), separated by an insulating barrier (I). 
	 Both devices are fabricated by electron beam lithography and three-angle shadow
	  evaporation~\cite{Shilpi}. The gate voltages $\rm V_{g, det}$  and $ \rm V_{g}$
	   are used to control the tunneling rates in the detector and source SET
	    respectively. 
	    The source bias voltage $V$ was set to 0 for sample A and was varied to change the tunneling rates in sample B.
	 }
\label{device_image}
\end{figure}

In this Letter we analyse random telegraph signal which switches between two levels.
We propose an 
approach, which has advantages if the detector is slow and
a threshold algorithm would produce too many errors to be corrected by the techniques mentioned above. For that purpose,
we use an analytical expression for the statistical
distribution of the output current of the detector recording 
random telegraph noise and having an arbitrary bandwidth. 
This distribution has been derived by Fitzhugh \cite{Fitzhugh} and used, for example, in the analysis
of switching rates between the charging states of superconducting double dot by Lambert {\it et al} \cite{Lambert}.  
We propose to reduce the effective bandwidth of the detector below its maximum value, set by hardware, 
by digital low pass filtering of the output signal. 
Varying the bandwidth in this way and comparing the cumulants of the resulting  distributions 
with corresponding analytical expressions, one can determine the switching rates. 
Our method should work even in the limit of a very slow detector, 
when the two current levels, between which the switching is happening,
cannot be reliably determined.

\begin{figure}
\includegraphics[width=\columnwidth]{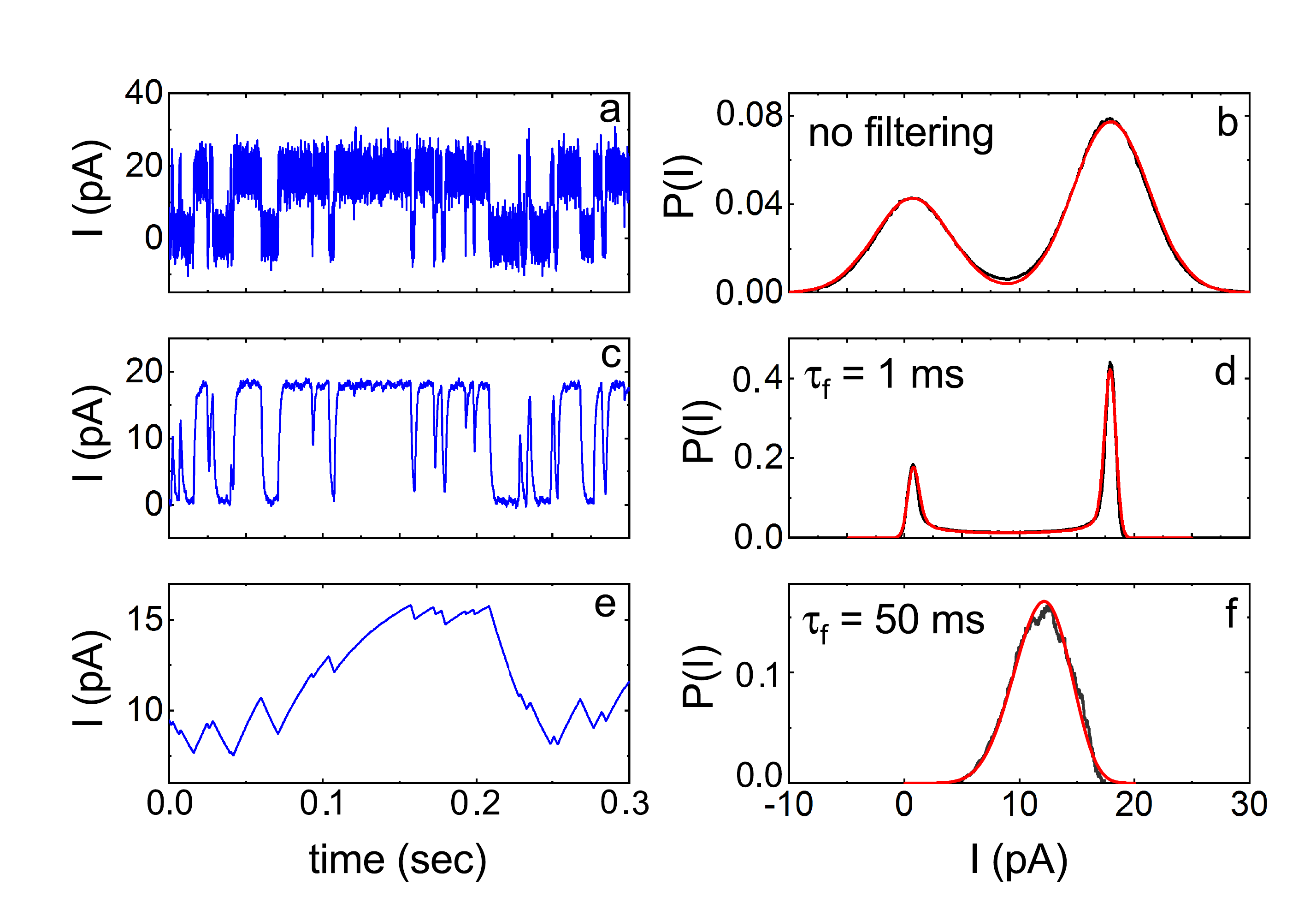}
\caption{
	\textbf{Sample A. }
	Time dependence of the current in the sample A before (a) and after the filtering (c,e),
and corresponding distributions (b,d,f).
In (b,d,f) black curves are experimental data,
and red curves are the fits with Eqs. (\ref{cal_P},\ref{P}). Fit parameters are indicated in the text and caption of Fig.~\ref{lifetime}.
	}
\label{histograms}
\end{figure}

In order to test the theory, we have studied random telegraph noise in a system of two capacitively coupled single-electron
transistors (SET)~\cite{Shilpi}. One of them is highly resistive and serves as a source
of telegraph noise caused by random switches between the two charge states
of the SET island~(turquoise in Fig.~\ref{device_image}). We denote these states as 
 0 (no extra electron on the island) and 1 (one extra electron on the island).
The second transistor with lower resistance is used as a detector~ (blue in Fig.~\ref{device_image}).
We voltage bias the detector~($V_{\rm b, det}$) above the Coulomb blockade threshold,
to have measurable current flowing through it,
and
tune the gate voltage~($V_{\rm g, det}$) to the most sensitive point. The detector current~($I$) monitors
the charging states of the noise source SET and switches between the two 
values $I_0$ and $I_1$ corresponding to the states $0$ and $1$. We denote the rate of the transition 
$0\to 1$ by 
$\gamma_\uparrow$, and the rate of 
the opposite transition, $1\to 0$, by  
$\gamma_\downarrow$.
On top of 
purely telegraph signal, $I_{\rm tel}(t)$, measurement setup adds the noise $\xi(t)$. 
Thus, the output current of the detector SET is
\begin{eqnarray}
I_d(t) = I_{\rm tel}(t) + \xi(t).
\label{Id}
\end{eqnarray}
We have fabricated two samples with similar design (Fig.~\ref{device_image}). For sample A
we have recorded 50 seconds long time traces of the detector current at zero source SET bias with the sampling
rate of $50$ kHz (time step  $\tau_0=20$ $\mu$s)
by digitizing the output of current preamplifier with an analog-to-digital converter. 
Part of such a  
trace is shown in Fig.~\ref{histograms}a.  
Sample A produces clear telegraph signal which can be analyzed by standard methods.
We use it as a reference to test the predictions of our model.
The detector of the second sample (sample B) has been embedded in a
radio-frequency resonant circuit and instead of the current we monitor the transmission $|s_{21}|^2$ through the circuit at $588 $ MHz.
This sample has also shown good  
telegraph signal with zero bias applied to 
the source SET~(Fig. \ref{histograms_2}a). We have deliberately applied higher bias to it
 in order to increase the transition rates
and to complicate the detection of  switching events~(Fig. \ref{histograms_2}c,e). 
For this sample we have recorded  
10 seconds long traces of the transmission coefficient $|s_{21}|^2$ with time step $\tau_0=0.1$ ms. 
Under these conditions, a typical 
histogram has only one peak  (Fig. \ref{histograms_2}d,f)  
and the threshold algorithm  
for determining the rates cannot be applied. We will demonstrate that 
our approach works also in this case. 

We will now briefly describe our theoretical model.
We assume that the output current of the detector SET (\ref{Id}) goes through a low pass filter
with the bandwidth $\tau_f^{-1}$, which transforms it as follows
\begin{eqnarray}
I(t)=\frac{1}{\tau_f}\int_{-\infty}^t dt' \, e^{-(t-t')/\tau_f} [I_{\rm tel}(t') + \xi(t')]. 
\label{It}
\end{eqnarray}
We have chosen this type of filtering because it emulates an $RC$ low pass filter, which is common in experiments,
and because it allows an exact solution of the problem. 
Our goal is to find the distribution of the filtered current $I(t)$.
As a first step, we solve the problem without noise and put $\xi(t)=0$.
We introduce two current distributions: one corresponding to the state 0 of the source SET, which we denote as ${\cal P}_0(t,I)$, 
and the second one, ${\cal P}_1(t,I)$, corresponding to the state 1. The evolution of these two distributions in time is described
by theory of stochastic jump processes (see, for example Ref.~\cite{BP} for details). 
The corresponding evolution equations read
\begin{eqnarray}
\partial_t {\cal P}_0 + \tau_f^{-1}\partial_I[(I_0-I) {\cal P}_0] 
&=& -\gamma_\uparrow {\cal P}_0 +\gamma_\downarrow  {\cal P}_1,
\nonumber\\
\partial_t {\cal P}_1 + \tau_f^{-1}\partial_I[(I_1-I) {\cal P}_1] 
&=& \gamma_\uparrow {\cal P}_0 - \gamma_\downarrow {\cal P}_1.
\label{FP}
\end{eqnarray}
The stationary solution of these equations can be found analytically \cite{Fitzhugh}. It reads
\begin{eqnarray}
{\cal P}_0(I) = \frac{I_1-I}{I_1-I_0}{\cal P}(I),\;\;
{\cal P}_1(I) = \frac{I-I_0}{I_1-I_0}{\cal P}(I),
\end{eqnarray}
where ${\cal P}(I)={\cal P}_0(I)+{\cal P}_1(I)$  reads \cite{Fitzhugh}
\begin{eqnarray}
{\cal P}(I) = \frac{\Gamma(\gamma_\Sigma\tau_f)}{\Gamma(\gamma_\uparrow\tau_f) \Gamma(\gamma_\downarrow\tau_f)}  
\frac{(I-I_0)^{\gamma_\uparrow\tau_f-1}(I_1-I)^{\gamma_\downarrow\tau_f-1}}{(I_1-I_0)^{\gamma_\Sigma\tau_f-1}}.
\label{cal_P}
\end{eqnarray}
Here $\Gamma (x)$ is the gamma function and $\gamma_\Sigma=\gamma_\uparrow+\gamma_\downarrow$. ${\cal P}(I)$ 
is the distribution of the filtered current (\ref{It}), which we are looking for. 
It has the form of 
a beta distribution, well known in statistics.  
The distribution  of Eq.~(\ref{cal_P})  
differs from the one derived for a filter with a sharp cutoff,
$I(t)=\frac{1}{\tau_f}\int_{t-\tau_f}^t dt' \,I_{\rm tel}(t')$, in Ref. \cite{Shilpi}, and it
does not have a universal form predicted in Ref. \cite{JS}. However, in the limit $\gamma_\Sigma\tau_f\gg 1$ 
all these distributions approach Gaussian form with the same parameters.
In the presence of noise the expression (\ref{cal_P}) should be convolved  
 with the distribution of the
filtered noise
$
W_\xi(I)=\left\langle\delta\left( I -  \frac{1}{\tau_f}\int_{-\infty}^t dt' \, e^{-(t-t')/\tau_f} \xi(t')\right)\right\rangle,
$
where $\langle\dots\rangle$ implies averaging over $\xi(t)$, and takes the form
\begin{eqnarray}
P(I) =\int_{I_0}^{I_1} dI'\, W_\xi(I-I')\,  {\cal P}(I').
\label{P}
\end{eqnarray}

The average current evaluated with the distribution (\ref{P}) does not depend on time,
\begin{eqnarray}
\langle I\rangle = (\gamma_\downarrow I_0+\gamma_\uparrow I_1)/\gamma_\Sigma.
\label{Iav}
\end{eqnarray}
The second cumulant ${\cal C}_2=\left\langle (I-\langle I \rangle)^2\right\rangle$ has the form
\begin{eqnarray}
{\cal C}_2(\tau_f) = \frac{\gamma_\uparrow\gamma_\downarrow(I_1-I_0)^2}{\gamma_\Sigma^2(1+\gamma_\Sigma\tau_f)} + \sigma_{\xi}^2(\tau_f),
\label{C2}
\end{eqnarray}
where $\sigma_{\xi}^2(\tau_f)$ is the variance of the filtered noise.
This parameter is expressed via the noise spectral power,
$S(\omega)=\int dt e^{i\omega t}\langle \xi(t)\xi(0)\rangle$, and reads
\begin{eqnarray}
\sigma_\xi^2(\tau_f) = \int_{\omega_{\min}}^\infty \frac{d\omega}{\pi} \frac{S(\omega)}{1+\omega^2\tau_f^2}.
\label{sigma0}
\end{eqnarray}
Typically $\xi(t)$ is the sum of white and $1/f$-noises, so that
$S(\omega)=2\eta + A/|\omega|$.
In this case one finds
\begin{eqnarray}
\sigma_{\xi}^2(\tau_f) = (\sigma_0^2-\sigma_\infty^2)({\tau_0}/{\tau_f}) + \sigma^2_\infty,
\label{sigma}
\end{eqnarray}
where $\sigma_0^2 = \sigma^2_\infty + \eta/\tau_0$ is the variance of the unfiltered current digitized with the time step $\tau_0$,
and $\sigma^2_\infty = ({A}/{2\pi})\ln\left[1+(\omega_{\min}\tau_f)^{-2}\right]$ 
is the contribution of $1/f$-noise. The latter is almost independent on $\tau_f$ until it exceeds $\omega_{\min}^{-1}$.
$\sigma^2_\infty$ may also include the contribution of
noise at 50 Hz due to pick-up from power lines.  

The third and the fourth cumulants of the current, 
${\cal C}_3=\left\langle (I-\langle I \rangle)^3\right\rangle$ and
${\cal C}_4=\left\langle (I-\langle I \rangle)^4\right\rangle - 3{\cal C}_2^2$ respectively, 
are not sensitive to the Gaussian white noise.
Normalized to their values at $\tau_f=0$, they read
\begin{eqnarray}
\frac{{\cal C}_3(\tau_f)}{{\cal C}_3(0)} &=&\frac{2}{(1+\gamma_\Sigma\tau_f)(2+\gamma_\Sigma\tau_f)} + \frac{{\cal C}_{3,\infty}}{{\cal C}_3(0)},
\label{C3}
\\
\frac{{\cal C}_4(\tau_f)}{{\cal C}_4(0)} &=&
\frac{(\gamma_\uparrow-\gamma_\downarrow)^2(1+\gamma_\Sigma\tau_f)-\gamma_\uparrow\gamma_\downarrow(2+\gamma_\Sigma\tau_f)}
{\gamma_\uparrow\gamma_\downarrow(1+\gamma_\Sigma\tau_f)^2(2+\gamma_\Sigma\tau_f)(3+\gamma_\Sigma\tau_f)} 
\nonumber\\ &&
+\, \frac{{\cal C}_{4,\infty}}{{\cal C}_4(0)}.
\label{C4}
\end{eqnarray}
The third cumulant of the unfiltered current reads
${\cal C}_3(0) = {\cal C}_{3,\infty} + {\gamma_\uparrow\gamma_\downarrow(\gamma_\downarrow-\gamma_\uparrow)(I_1-I_0)^3}/{\gamma_\Sigma^3}$.
The expression for ${\cal C}_4(0)$ is rather long, and we skip it for simplicity.
The long time limiting values ${\cal C}_{3,\infty}$ and ${\cal C}_{4,\infty}$ account for the effect of non-Gaussian $1/f$-noise, 
and weakly depend on $\tau_f$. Importantly, the $\tau_f$-dependent parts of the normalized cumulants
(first terms in r.h.s. of Eqs. (\ref{C3},\ref{C4}))
do not contain the current levels $I_0$ and $I_1$, which are impossible to determine for a slow
detector with a single peak current distribution (see e.g. Fig. 2f, Fig. 4d). 
Equations (\ref{C3},\ref{C4}) provide the basis for finding the rates in this case.

Having developed the theoretical model, we test it with the data recorded from the reference sample A.
For this sample the distribution of the unfiltered current is well described by the
sum of the two Gaussian peaks centered around the currents $I_0 = 0.71$ pA, $I_1=17.97$ pA, and
having the width  $\sigma_0=3.32$ pA   (Fig. \ref{histograms}b).

\begin{figure}
\begin{tabular}{cc}
\includegraphics[width=0.5\columnwidth]{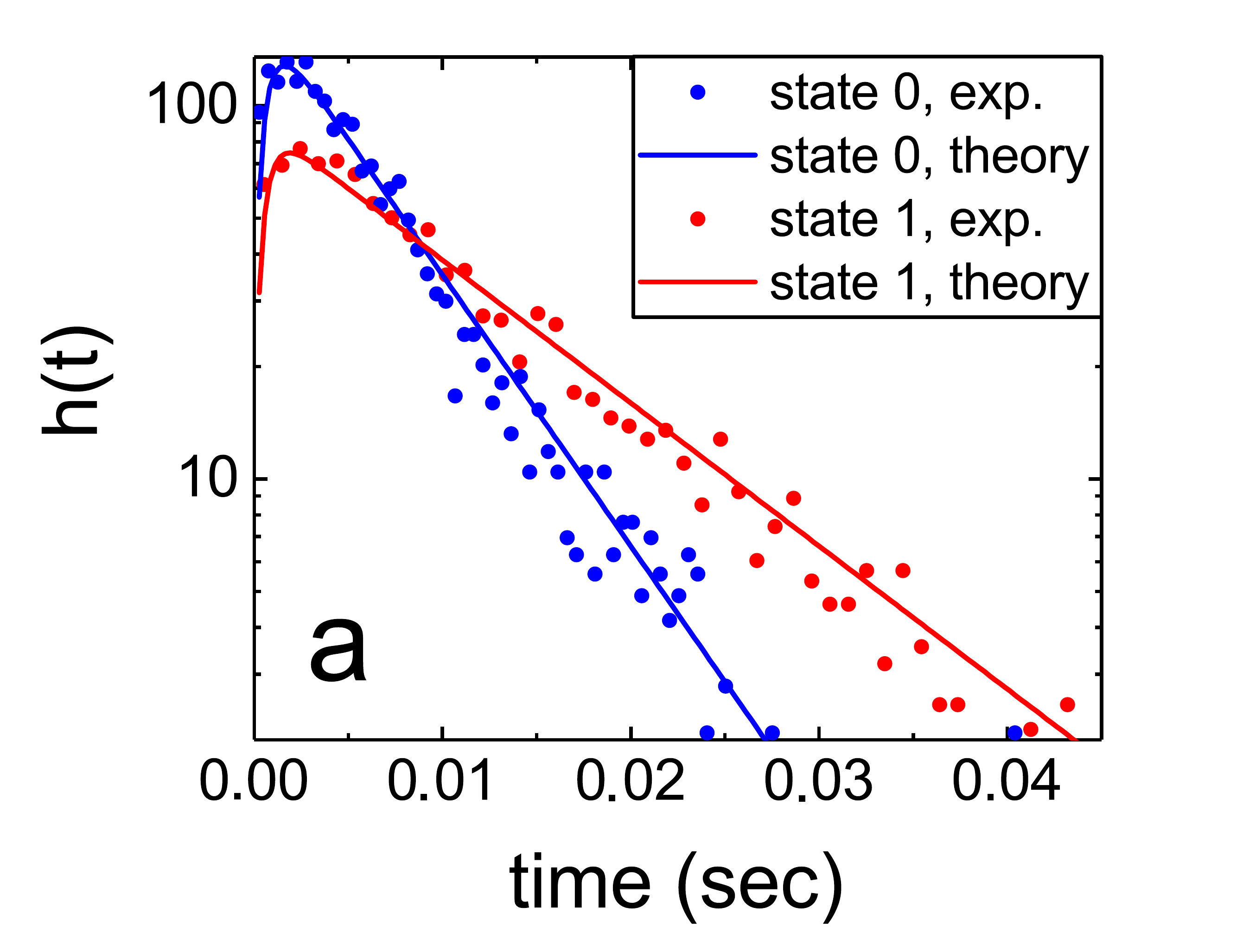} &  \includegraphics[width=0.5\columnwidth]{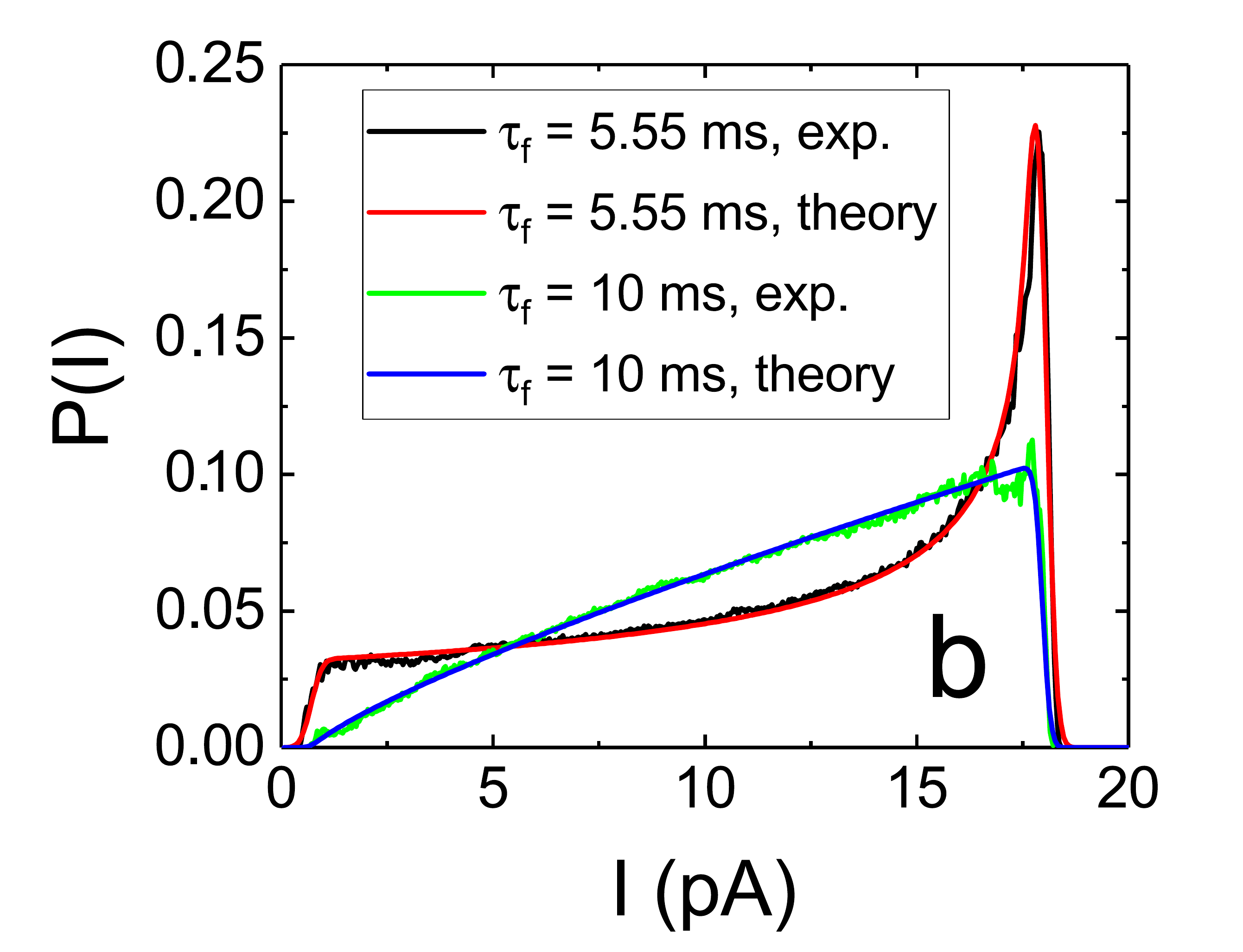}
\end{tabular}
\caption{ \textbf{Calculation of rates.}
	(a) Distributions of the life-times of the states 0 and 1 of 
	 sample A extracted 
	from the current time trace filtered with $\tau_f=1$ ms (symbols) 
	based on conventional threshold detection.  
	Solid lines show Eq. (\ref{survival}) with the rates $\gamma_\uparrow=180$ Hz, $\gamma_\downarrow=100$ Hz.
	(b) Distributions of the detector current at  $\tau_f = \gamma_\uparrow^{-1} = 1/180 $ s and $\tau_f = \gamma_\downarrow^{-1} = 1/100$ s. 
	Fits are based on Eqs. (\ref{cal_P1}) and (\ref{P}). 
	}
\label{lifetime}
\end{figure}

\begin{figure}
\includegraphics[width=\columnwidth]{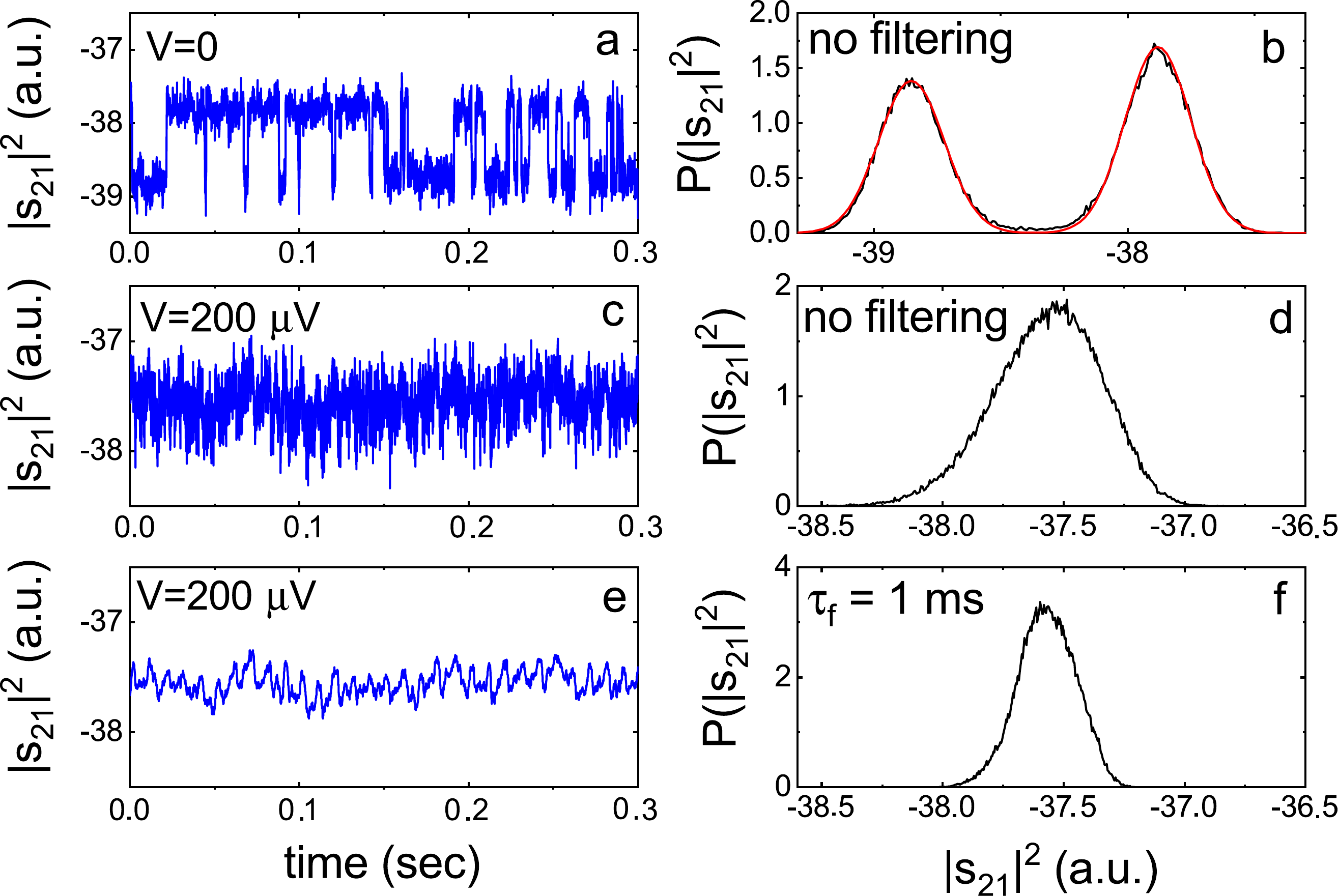}
\caption{ 	\textbf{Sample B.}
	Time traces of $|s_{21}|^2$ at zero bias 
		and dimensionless gate voltage (here $C_g$ is the gate capacitance) 
		$C_gV_g/e = 0.48$, which is close to the
		maximum of conductance peak,
	with the rates $\gamma_\uparrow=220$ Hz, $\gamma_\downarrow= 180$ Hz (a); 
	and at $V=200$ $\mu$V, $C_gV_g/e=0.42$  (c,e).
    No filtering was applied to the traces (a) and (c), while the trace (e) has been
    filtered with $\tau_f=1$ ms. The corresponding distributions are shown in (b,d,f).
    In (d,f) theory fits are absent because the values of the transmission coefficient, 
    between which the switching is happening, are not precisely known at $V=200$ $\mu$V.  
    Red curve in (b) is a fit with two Gaussian peaks.   
	}
	
\label{histograms_2}
\end{figure}

Next, we numerically generate filtered current time traces (\ref{It}) for different values of $\tau_f$.  
Technically, we first perform discrete Fourier transformation of the unfiltered current
generating a sequence of Fourier components $\tilde I_k = \sum_j I_j e^{-2\pi i(i-1)(k-1)/N}/\sqrt{N}$,
where $N$ is the total number of points $I_j$ in the data set. Afterwards we multiply them by the filtering function 
\begin{eqnarray}
F_k = \frac{1-e^{-\tau_0/\tau_f}}{1-e^{-\tau_0/\tau_f} e^{-2\pi i(k-1)/N}},
\label{Fk}
\end{eqnarray}
which is the discrete Fourier transform of the
exponent appearing in Eq.~(\ref{It}). Finally, we
apply the inverse Fourier transformation.
Two filtered time traces generated in this way are shown in Figs.~\ref{histograms}c and \ref{histograms}e, and one more for sample B in Fig.~\ref{histograms_2}e.   
We have generated a series of current distributions with different filtering times and fitted them
with Eqs. (\ref{cal_P},\ref{P}).
We have achieved the best fit of the data with the switching rates $\gamma_\uparrow=180$ Hz, $\gamma_\downarrow=100$ Hz, and
assuming Gaussian form
of the noise current distribution $W(I)=\exp(-I^2/2\sigma^2)/\sqrt{2\pi}\sigma$, where $\sigma$ is defined by 
Eq.~(\ref{sigma})
with $\sigma_\infty=1.04$ pA.
With these parameters we have fitted current histograms in the wide range of averaging times $0<\tau_f<0.5$ sec. 
Examples of such fits are shown in   
Figs. \ref{histograms}d, \ref{histograms}f and \ref{lifetime}b. 

For comparison, we have also determined the switching rates in a  
usual way.  
For that purpose we have
plotted the distributions $h_0(t)$ and $h_1(t)$  of the lifetimes
of the states 0 and 1 respectively. 
The lifetimes have been obtained from the current trace filtered with $\tau_f=1$ ms, for which 
we have applied a standard threshold detection algorithm with the threshold between the states 0 and 1 placed at $(I_0+I_1)/2$.
In Fig. \ref{lifetime}a, we have compared the resulting distributions with the formula derived in Ref. \cite{NA},
\begin{eqnarray}
h_{0,1}=\frac{2\gamma_{\uparrow,\downarrow}\gamma_{\rm det} e^{-\lambda t/2}
\sinh\left( \sqrt{\lambda^2-4\gamma_{\uparrow,\downarrow}\gamma_{\rm det}}\,t/2\right)}
{\sqrt{\lambda^2-4\gamma_{\uparrow,\downarrow}\gamma_{\rm det}}}.
\label{survival}
\end{eqnarray}
Here $\lambda=\gamma_\Sigma+\gamma_{\rm det}$ and $\gamma_{\rm det}=(\tau_f\ln 2)^{-1}=1443$ Hz
is the effective detector bandwidth. 
We have found very good agreement between theory and experiment with the same values of the switching rates as before,
see Fig. \ref{lifetime}a.

The shape of the current distribution changes with the 
filtering time $\tau_f$ in a way illustrated in Fig. \ref{histograms}. 
At short $\tau_f$ the white noise is suppressed and the two peaks in the distribution become sharper. 
With growing $\tau_f$ the two peaks disappear one after another
exactly at times $\tau_f=\gamma_\uparrow^{-1}$ and $\tau_f=\gamma_\downarrow^{-1}$.
At $\tau_f=\gamma_\uparrow^{-1}$ 
 the current distribution (\ref{cal_P}) takes a particularly simple form 
\begin{eqnarray}
\label{cal_P1}
{\cal P}(\tau_f=\gamma_\uparrow^{-1},I) = \frac{\gamma_\downarrow(I_1-I)^{\gamma_\downarrow/\gamma_\uparrow-1}} 
{\gamma_\uparrow(I_1-I_0)^{\gamma_\downarrow/\gamma_\uparrow}}
\label{P1}
\end{eqnarray}
shown in Fig. \ref{lifetime}b.
The distribution at $\tau_f=\gamma_\downarrow^{-1}$ 
is given by the same formula with the interchanged rates $\gamma_\uparrow\leftrightarrow\gamma_\downarrow$.

\begin{figure}
\includegraphics[width= \columnwidth]{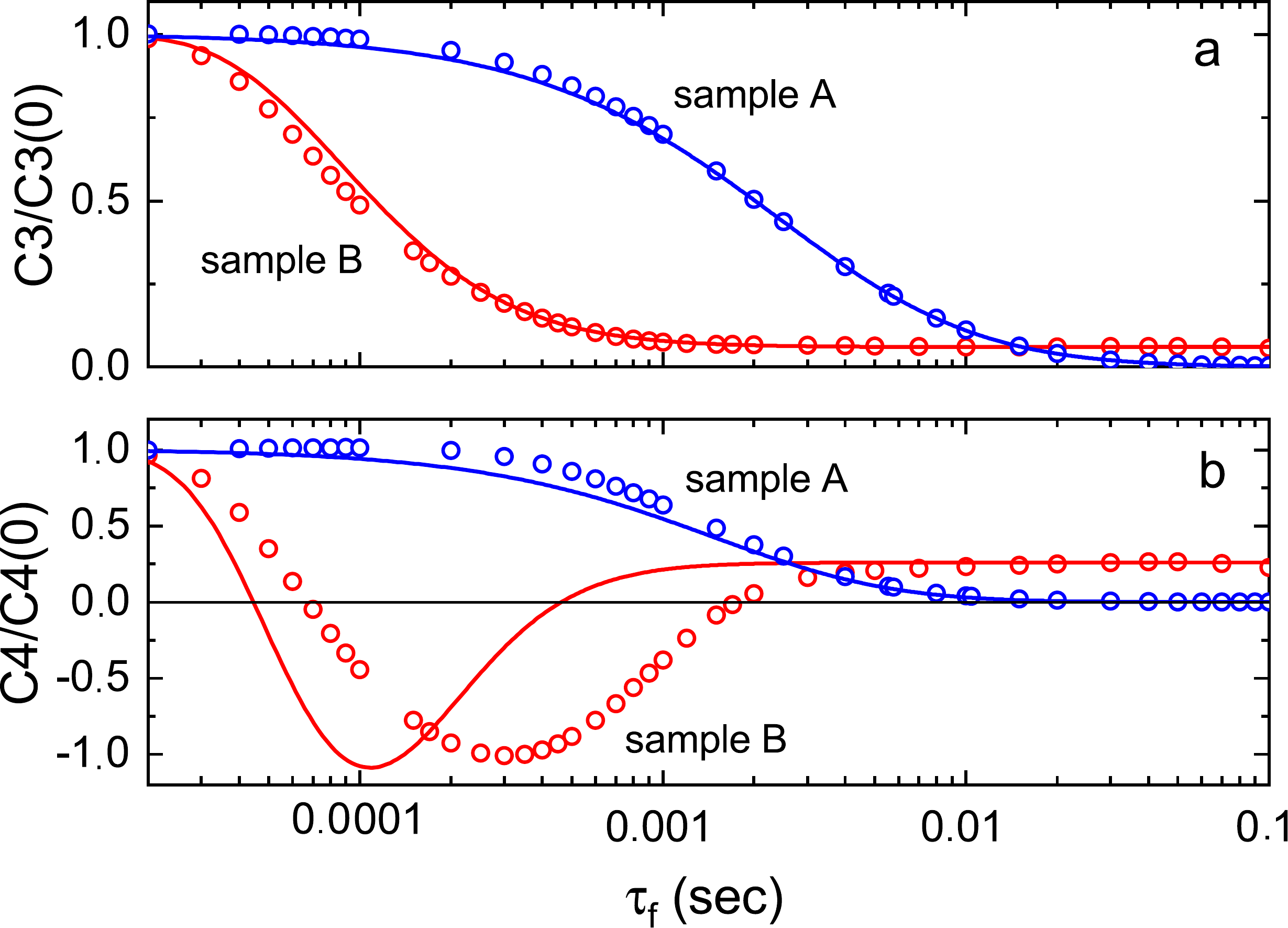}  
\caption{
	\textbf{Using cumulants as rate calculator.}
	Normalized third (a) and fourth (b) cumulants of the distributions of the current for sample A (blue) 
	and of the transmission coefficient for sample B (red). Symbols are experimental data, solid lines  theory fits  with Eqs. (\ref{C3},\ref{C4}). 
	For sample A we used the parameters reported in the text and ${\cal C}_{3,\infty}={\cal C}_{4,\infty}=0$. 
	For  sample B we have found ${\cal C}_{3,\infty}=0.06 {\cal C}_3(0)$ and ${\cal C}_{4,\infty}=0.24 {\cal C}_4(0)$.
	For both samples the third cumulants are negative, ${\cal C}_3(\tau_f)<0$, implying $\gamma_\uparrow>\gamma_\downarrow$. 
	}
\label{C3C4}
\end{figure}

\begin{figure}
\includegraphics[width=\columnwidth]{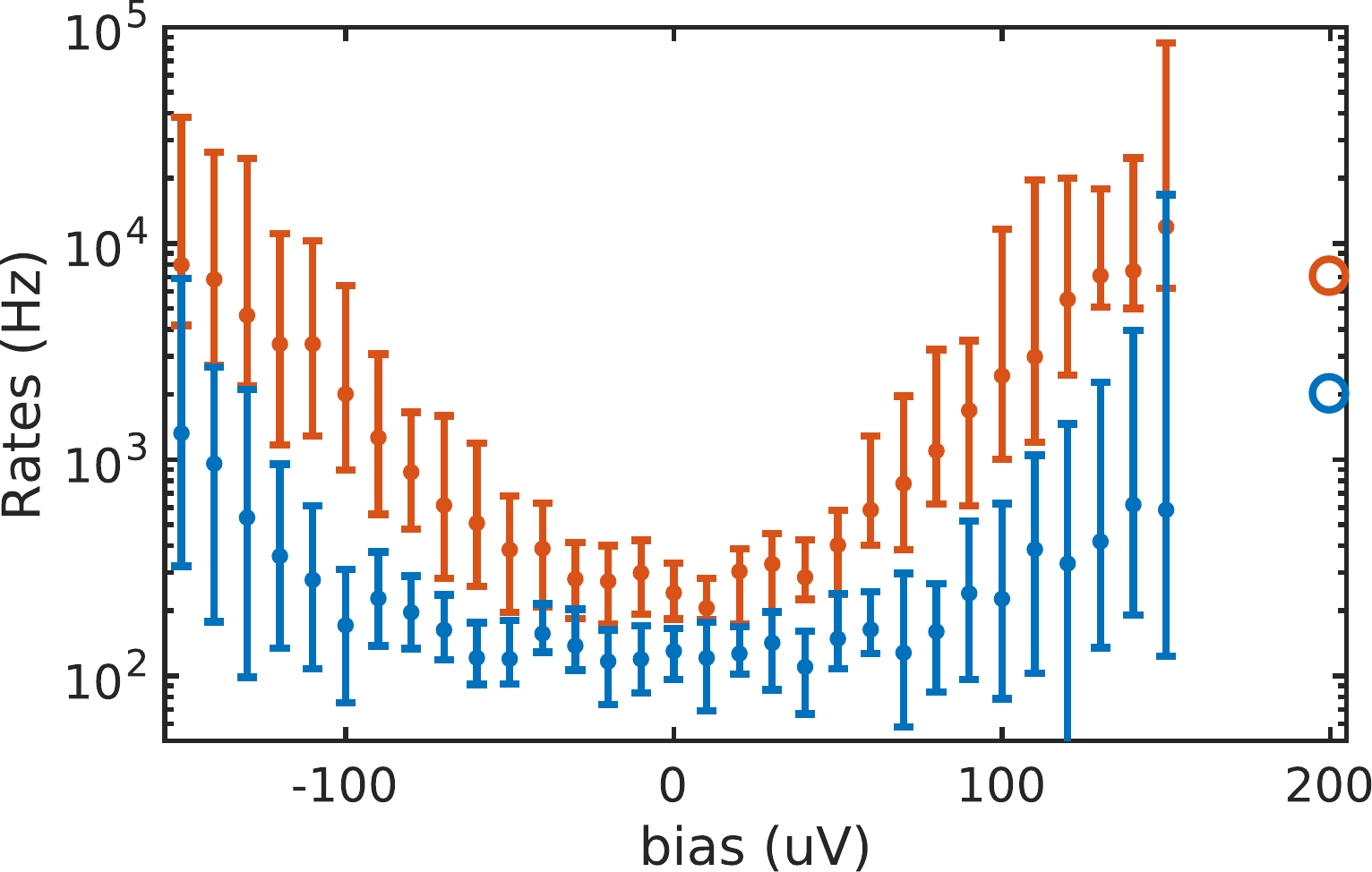} 
\caption{
	\textbf{Switching rates as a function of source SET bias voltage for sample B.}
	Bias dependence of the tunneling rates $\gamma_\uparrow$ (red circles) and $\gamma_\downarrow$ (blue circles) for the sample B
    for a fixed value of the gate voltage, $C_gV_g/e=0.42$,  applied to the SET. 
    The rates are extracted with traditional threshold-based methods, with~error bars
	originating from measurement limitations, like small drift in gate voltage between different bias voltages and error in choice of effective detector bandwidth while using~\cite{NA} for rate calculation. 
    Open circles are the rates calculated with the method presented here.
	}
\label{rates_bias}
\end{figure}

Next, we apply our model to  
sample B, replacing the current by transmission
coefficient, $I\to |s_{21}|^2$. The distribution of the latter has a form of a single skewed peak,
which does not split into two peaks even after filtering. 
The shape of the peak resembles the distribution
of the heavily filtered current of sample A, cf. Figs. \ref{histograms}f and \ref{histograms_2}d.
Obviously, standard threshold detection of the 
switching events is not possible in this case.
In order to estimate the switching rates  we
evaluate the cumulants ${\cal C}_3$ and ${\cal C}_4$, plot them as 
functions of the time $\tau_f$ (Fig. \ref{C3C4}) and fit the result with Eqs. (\ref{C3},\ref{C4}). 
The data from sample A (blue) can be reasonably well fitted with the same parameters as before, which once
again confirms the validity of our model. Fitting the third cumulant for the sample B (red), 
we have determined the total switching rate in this sample, $\gamma_\Sigma=9\pm 2$ kHz.
Next, fitting the fourth cumulant we  
determine the two rates separately, $\gamma_\uparrow=7$ kHz and $\gamma_\downarrow=2$ kHz.
Strong low frequency noise and finite value of the time step limit the accuracy of the result, especially for the fourth cumulant.
We have verified that these values of the rates are consistent 
with the ones obtained by extrapolation from the lower bias regime, 
where the standard threshold-based methods are still applicable, see Fig. \ref{rates_bias}.
To account for the finite sampling rate in the simplest approximation  we have replaced $\tau_f\to \tau_f^{\rm eff}=\tau_0/(e^{\tau_0/\tau_f}-1)$ in
Eqs. (\ref{C3},\ref{C4}). This replacement is justified by a low-frequency expansion of the function (\ref{Fk}) and by fitting
it to the form $1/(1-i\omega \tau_f^{\rm eff})$, corresponding to  exponential decay in the time domain.
  
In conclusion, we have proposed a  
method of determining the switching rates 
of a random telegraph noise, which is based on an analytical expression for the distribution 
of the filtered signal, and which should work even if the detector is slow.
We have confirmed the validity of the model 
applying it to the telegraph signal generated by a single-electron transistor, for which
the switching rates can be determined by a conventional 
threshold algorithm. Subsequently, we have demonstrated the efficiency of the method
applying it to the device with  
a slow detector, for which the current threshold between the two states cannot be defined.
We believe that the formalism presented here can be extended to multi-level telegraph noise.

We acknowledge helpful discussions with Ivan Khaymovich.
We acknowledge the provision of facilities by Aalto
University at OtaNano – Micronova Nanofabrication
Centre. We thank Matthias Meschke for technical assistance. 
  This work is partially supported by Academy of
Finland, Project Nos. 284594, 272218, and 275167 (S. S., 
D. S. G.,  J. T. P., and J. P. P.).


\begin{thebibliography}{99}

\bibitem{Weissman} M.~B. Weissman, Rev. Mod. Phys. {\bf 60}, 537 (1988).
\bibitem{KU}  M. J. Kirton and M. J. Uren, Adv. Phys. {\bf 38}, 368 (1989).
\bibitem{MOSFET} Z.~Li, M.~Sotto, F.~Liu, M.~K.~Husain, H.~Yoshimoto,
Y.~Sasago, D.~Hisamoto, I.~Tomita, Y.~Tsuchiya, and S.~Saito, Sci. Rep. {\bf 8}, 250 (2018).
\bibitem{NA1} O. Naaman and J. Aumentado, Phys. Rev. B {\bf 73}, 172504 (2006).
\bibitem{SET} T. M. Buehler, D. J. Reilly, R. P. Starrett, V. C. Chan, A. R. Hamilton, A. S. Dzurak, and R. G. Clark,
J. Appl. Phys. {\bf 96}, 6827 (2004).
\bibitem{Hofheinz} M. Hofheinz, X. Jehl, M. Sanquer, G. Molas, M. Vinet, and S. Deleonibus, Eur. Phys. J. B {\bf 54}, 299 (2006).
\bibitem{qubit} M.~M\"ott\"onen, R.~de~Sousa, J.~Zhang, and K.~B.~Whaley, Phys. Rev. A {\bf 73}, 022332 (2006).
\bibitem{Delsing} M. Shaw, R. Lutchyn, P. Delsing, and P. Echternach, Phys. Rev. B {\bf 78}, 024503 (2008).
\bibitem{Ustinov} G.~J. Grabovskij, T. Peichl, J. Lisenfeld, G. Weiss, A.~V.~Ustinov, Science {\bf 338}, 232 (2012).
\bibitem{NA} O. Naaman and J. Aumentado, Phys. Rev. Lett. {\bf 96}, 100201 (2006).
\bibitem{YYJ} Y. Yuzhelevski, M. Yuzhelevski, and G. Jung, Rev. Sci. Instr. {\bf 71}, 1681 (2000).
\bibitem{Martinez} J. Martin-Martinez, J. Diaz, R. Rodriguez, M. Nafria, and X. Aymerich, IEEE Electron Device Letters {\bf 35}, 479 (2014). 
\bibitem{PhysRevB79035314} B. K\"ung, O. Pf{\"a}ffli, S. Gustavsson, T. Ihn, K. Ensslin, M.~Reinwald and  W. Wegscheider, Phys. Rev. B {\bf 79}, 035314 (2009).
\bibitem{Prance} J.~R. Prance, B.~J. Van Bael, C.~B. Simmons, D.~E. Savage, M.~G. Lagally, M. Friesen, S.~N. Coppersmith and M.~A. Eriksson,
Nanotechnology {\bf 26}, 215201 (2015).
\bibitem{Ensslin} S. Gustavsson, R. Leturcq, M. Studer, I. Shorubalko, T. Ihn, K. Ensslin, D.C. Driscoll, and A.C. Gossard, Surface Science Reports {\bf 64}, 191 (2009).
\bibitem{Haug} N. Ubbelohde, C. Fricke, C. Flindt, F. Hohls, and R. J. Haug, Nat. Commun. {\bf 3} 612 (2012).
\bibitem{Slichter} D. H. Slichter, R. Vijay, S. J. Weber, S. Boutin, M. Boissonneault, J. M. Gambetta, A. Blais, and I. Siddiqi, Phys. Rev. Lett. {\bf 109}, 153601 (2012).
\bibitem{Rabiner} L. R. Rabiner,  Proceedings of the IEEE {\bf 77}, 257 (1989).
\bibitem{Awano} H. Awano, H. Tsutsui, H. Ochi and T. Sato,  International Symposium on Quality Electronic Design (ISQED), 14th International Symposium on. IEEE, (2013).
\bibitem{Puglisi} F. M. Puglisi, P. Pavan, ECTI Transactions on Electrical Engineering, Electronics, and Communications,  {\bf 12}, 24 (2014).
\bibitem{Giusi} G. Giusi, F. Crupi and C. Pace, Rev. Sci. Instr. {\bf 79}, 024701 (2008).
\bibitem{Fitzhugh} R. Fitzhugh, Math. Biosci. {\bf 64}, 75 (1983).
\bibitem{Lambert} N. J. Lambert, A. A. Esmail, F. A. Pollock, M. Edwards, B. W. Lovett, and A. J. Ferguson, Phys. Rev. B {\bf 95}, 235413 (2017).
\bibitem{BP} H. P. Breuer and F. Petruccione, {\it The theory of open quantum systems}, Oxford University Press (2007).
\bibitem{Shilpi} S. Singh, J. T. Peltonen, I. M. Khaymovich, J. V. Koski, C. Flindt, and J. P. Pekola, Phys. Rev. B {\bf 94}, 241407(R) (2016).
\bibitem{JS} A. N. Jordan and E. V. Sukhorukov, Phys. Rev. Lett. {\bf 93}, 260604 (2004).
\end{thebibliography}
\end{document}